 \documentclass[pmlr,twocolumn]{jmlr} % W&CP article

\usepackage{booktabs}
\usepackage[load-configurations=version-1]{siunitx} % newer version
\newcommand{\xl}[1]{{\color{black} #1}}

\theorembodyfont{\upshape}
\theoremheaderfont{\scshape}
\theorempostheader{:}
\theoremsep{\newline}

\usepackage{makecell}
\graphicspath{{figures/}}

\jmlrvolume{ML4H Extended Abstract Arxiv Index}
\jmlryear{2020}
\jmlrsubmitted{2020}
\jmlrpublished{}
\jmlrworkshop{Machine Learning for Health (ML4H) 2020} % W&CP title

\title[staDRIP]{A stability-driven protocol for drug response interpretable prediction (staDRIP)}

\author{%
\Name{Xiao Li}\nametag{\thanks{The first two authors contributed equally to this paper.}}\Email{sxli@berkeley.edu}\\
\emph{\textmd{University of California, Berkeley}}\\
\Name{Tiffany M. Tang} \Email{tiffany.tang@berkeley.edu}\\
\emph{\textmd{University of California, Berkeley}}\\
\Name{Xuewei Wang} \Email{Wang.Xuewei@mayo.edu}\\
\emph{\textmd{Mayo Clinic}}\\
\Name{Jean-Pierre A. Kocher} \Email{Kocher.JeanPierre@mayo.edu}\\
\emph{\textmd{Mayo Clinic}}\\
\Name{Bin Yu} \Email{binyu@berkeley.edu}\\
\emph{\textmd{University of California, Berkeley}}
}

\begin{document}

\maketitle

\begin{abstract}
Modern cancer -omics and pharmacological data hold great promise in precision cancer medicine for developing individualized patient treatments. However, high heterogeneity and noise in such data pose challenges for predicting the response of cancer cell lines to therapeutic drugs accurately. As a result, arbitrary human judgment calls are rampant throughout the predictive modeling pipeline. In this work, we develop a transparent stability-driven pipeline for drug response interpretable predictions, or staDRIP, which builds upon the PCS framework for veridical data science \citep{yu2020veridical} and mitigates the impact of human judgment calls. 
%
%The PCS framework was originally motivated by extensive interdisciplinary research, including those in biology and genomics, and it has since demonstrated its value in various scientific knowledge discovery problems in healthcare. 
%
Here we use the PCS framework for the first time in cancer research to extract proteins and genes that are important in predicting the drug responses and stable across appropriate data and model perturbations. Out of the 24 most stable proteins we identified using data from the Cancer Cell Line Encyclopedia (CCLE), 18 have been associated with the drug response or identified as a known or possible drug target in previous literature, demonstrating the utility of our stability-driven pipeline for knowledge discovery in cancer drug response prediction modeling.
\end{abstract}
%\begin{keywords}
%List of keywords
%\end{keywords}

\setcitestyle{numbers}

%\vspace{-5mm}

\section{Introduction}
\label{sec:intro}

A critical goal in precision medicine oncology revolves around predicting a patient's response to therapeutic drugs given the patient's unique molecular profile \citep{rubin2015health, kohane2015ten}. Accurate personalized drug response predictions can immediately shed light on therapies that are likely to be ineffective or toxic and aid clinicians in deciding the most promising treatment for their patients \citep{azuaje2016computational}. Moreover, interpreting these drug response prediction models can help to improve recommendations of compounds and target genes to prioritize in future preclinical research \citep{caponigro2011advances}. 

While several community-wide, public efforts \citep{barretina2012cancer, costello2014} and many other works 
%(\cite{iorio2016landscape, jang2014systematic, menden2013machine, yang2012genomics, ali2019machine, guvencc2019improving}, and references therein) 
have made progress towards improving the predictive accuracy of drug response predictions, identifying the important disease signatures (i.e., proteins, genes, and other biomarkers) that drive the drug response prediction models has received less attention. To date, previous works 
%involving the selection of response-predictive biomarkers from supervised machine learning algorithms 
have typically focused on feature selection within one specific model such as elastic nets \citep{jang2014systematic, barretina2012cancer} %ridge regression \citep{geeleher2017discovering},
and random forest \citep{riddick2011predicting}. However, because molecular profiling data is often heterogeneous, noisy, and high-dimensional, these results are highly sensitive to modeling decisions made by humans including the type of model, the amount of training data, and the choice of algorithm. %This instability and lack of understanding of predictive disease signatures is a major impediment to the future of precision medicine and drug response modeling. % given the heavy reliance on hypothesis-driven studies in the past. For instance, hypothesis-driven studies have been particularly important for discovering new therapies to combat cancer, which remains one of the deadliest diseases in the world despite the advances in chemotherapy and immunotherapy \citep{}.
%The fundamental challenge here in drug response prediction modeling stems from the existing data itself. Because cell lines, patients, and cancers are all incredibly heterogeneous and low signal-to-noise, prediction models trained on this data often lead to poor generalization performance, and extracting biological knowledge from these models is questionable at best. To make matters worse, due to feasibility and cost, molecular profiling data typically has small sample sizes (or number of observations) on the order of tens to hundreds but an enormous number of measured biomarkers (or features) on the order of tens of thousands. This data limitation compounds the poor generalizability issue and more importantly, results in little statistical power for detecting the important disease signatures that drive the prediction models.
%\vspace{-1mm}

%In this work, we develop a stability-driven pipeline for drug response interpretable prediction called staDRIP, which aims to detect stable, interpretable, and predictive \makebox{-omic} signatures that drive a cell's response to a specific drug.

In this work, we focus on this goal of detecting stable, interpretable, and predictive -omic signatures that drive a cell line's drug response. To overcome the aforementioned challenges, we develop a transparent stability-driven pipeline for drug response interpretable prediction called staDRIP that is rooted in the PCS framework for veridical data science \citep{yu2020veridical}. At its core, the PCS framework 
%synthesizes, streamlines, and expands on ideas and best practices in machine learning and statistics, 
builds its foundation on three principles: \textit{predictability} as a reality check, \textit{computability} as an important consideration in algorithmic design and data collection, and \textit{stability} as an overarching principle and minimal requirement for scientific knowledge extraction. These principles were motivated by extensive interdisciplinary research such as \cite{wu2016stability}, which analyzed the gap-gene network of \emph{Drosophila}, and \cite{basu2018}, which discovered stable transcription factor interactions in \emph{Drosophila} embryos. Since its conception, the PCS framework has further demonstrated a strong track record of driving many scientific discoveries including novel gene-gene interactions for the red-hair phenotype \citep{behr2020epitree} and clinically-interpretable subgroups in a randomized drug trial \citep{dwivedi2020stable}. 

Here, using integrative -omics and drug response data from the Cancer Cell Line Encyclopedia (CCLE)\footnote{The CCLE is one of the most comprehensive public databases for developing detailed genetic and pharmacologic characterizations of human cancer cell lines. After preprocessing (see Appendix~\ref{sec:data}), we arrive at a panel of 370 human cancer cell lines that have both high-throughput molecular profiling of RNASeq gene expression (5000 genes), microRNA expression (734 miRNAs), DNA methylation (4000 transcription start sites), and protein expression (214 proteins) as well as pharmacological data for 24 anticancer drugs.} \citep{barretina2012cancer}, we employ the PCS framework to develop staDRIP and provide extensive \href{https://github.com/Yu-Group/staDRIP}{documentation} of our modeling choices to arrive at stable biological discoveries of proteins and genes that are predictive of cancer drug responses. Unlike previous works whose results depend heavily on human decisions, staDRIP finds predictive \makebox{-omic} features that are stable across various models and data perturbations, thus mitigating the impact of human judgment calls. We further show that 18 of the top 24 -omic features identified by %the PCS inference in
staDRIP have been previously implicated in the scientific literature, and in doing so, hint at novel candidates for future preclinical research.

\section{Results} 
\label{sec:results}

\subsection{Prediction Accuracy} \label{sec:pred_results}

Building on the PCS framework, staDRIP first uses predictive accuracy as a reality check to filter out models that are poor fits for the observed data before turning to our primary goal of identifying important biomarkers for drug response prediction. \xl{Specifically, we define drug response as the area under the fitted dose-response curve of growth inhibition (see Appendix \ref{sec:data} for details)}.
For each of the available 24 anticancer drugs, we divide the data into a 50-25-25\% training-validation-test split and use the training data to fit (1) an elastic net tuned with cross-validation (CV), which has been widely used and advocated by previous studies \citep{barretina2012cancer, jang2014systematic}, (2) Lasso tuned with CV, (3) Lasso tuned with ESCV, \xl{an alternative CV metric that incorporates stability to yield more stable estimation properties with minimal loss of accuracy}
\citep{lim2016estimation}, (4) Gaussian kernel ridge regression, and (5) random forest to predict the drug response given the miRNA, RNASeq, methylation, and protein expression profiles separately. We also fit several data integration methods including concatenated versions of the aforementioned methods, 
%(i.e., applying the Lasso, kernel ridge, or RF to the data set where we concatenate all four molecular profiles), 
the recently proposed X-shaped Variational Autoencoder (X-VAE) \citep{simidjievski2019variational}, and the winner of the DREAM 7 challenge,\footnote{The DREAM 7 challenge %in 2012 
was a public competition 
%with 44 team submissions, 
where teams were tasked to integrate multiple –omics measurements and predict drug sensitivity in cancer cell lines.} the Bayesian multi-task multiple kernel learning method (BMTMKL) \citep{costello2014}.

For each of these fits, we report in Table~\ref{tab:pred-res} \xl{in Appendix \ref{apd:pred_res}} the average validation accuracy across all 24 drugs as measured by the $R^2$ value and the WPC-index, a weighted probabilistic concordance index, which has been used in previous studies and measures how well the predicted rankings agree with the true responses \citep{costello2014}. From Table~\ref{tab:pred-res}, we see that kernel ridge regression trained only on the RNASeq data yields the best predictive performance.
%giving the highest validation $R^2$ and WPC-index. 
%that for each method, the RNASeq profile contains the most predictive power out of the available profiling data. 
%and even outperforms the data integration methods.
However, considering that our primary goal is not purely prediction, the differences between model prediction accuracies shown in Table~\ref{tab:pred-res} are relatively small from a practical viewpoint. In our inferential procedure discussed next, we will see that leveraging the stability across these methods with similar predictive accuracies is key to our staDRIP pipeline for identifying genes and proteins that are stable predictive features underlying the drug response models.

%However, rather than relying solely on the validation prediction accuracy, we, in accordance with the PCS framework, emphasize the need for prediction stability to mitigate the poor generalizability issues suffered by existing drug response prediction pipelines. Toward this end, we repeatedly fit each method under consideration on 100 bootstrap replicates of the training data to evaluate its prediction stability on the validation set and report ... in Table~\ref{tab:pred-stab}.
% \vspace{-1mm}

Nonetheless, for completeness, we report the test accuracy from the best model, the RNASeq-based kernel ridge regression, to have an $R^2$ ($\pm 1$SD) of 0.204 ($\pm$ 0.038) and WPC-index of 0.620 ($\pm$ 0.0075) across the 24 drugs. For brevity, we 
%have only included the prediction results, averaged across all drugs, here and 
leave more detailed individual drug-level prediction results to the Appendix.

\subsection{Identifying predictive -omic features with PCS inference} \label{sec:interpret}

\begin{table*}[ht]
\scriptsize
\centering
\caption{Most stable protein associated with each drug, as identified by staDRIP, along with literature that supports the association between the protein and drug sensitivity. \label{tab:pubmedliterature}}
%\vspace{-2mm}
{\begin{tabular}{@{}ccc|ccc@{}}\toprule
\textbf{Drug} & \textbf{Protein} & \textbf{Supporting Literature} & \textbf{Drug} & \textbf{Protein} & \textbf{Supporting Literature} \\ \midrule
17-AAG &Bax&\citet{he2013hsp90}&PD-0332991&\makecell{Bcl-2}&\citet{chen2017dual}\\ \hline
%AEW541&\makecell{PTEN}&\makecell{\cite{patel2014pten}\\\cite{isebaert2011insulin}}\\ \hline
AEW541&Akt&\citet{attias2011insulin}&PF2341066&c-Met&\citet{camidge2014efficacy}\\ \hline
AZD0530&p38
&\citet{yang2010effect}&PHA-665752&MEK1&--\\ \hline
AZD6244&\makecell{PI3K-p85}&\citet{balmanno2009intrinsic}&PLX4720&\makecell{MEK1}&\citet{emery2009mek1}\\ \hline
Erlotinib&EGFR&\citet{mcdermott2007identification}&Paclitaxel&Src&\citet{le2011src}\\ \hline
Irinotecan&\makecell{MDMX\_MDM4}&\citet{ling2014fl118}&Panobinostat&VEGFR2&\citet{strickler2012phase}\\ \hline
L-685458&YAP&--&RAF265&PI3K-p85&\citet{mordant2010dependence}\\ \hline
LBW242&ASNS&--&Sorafenib&Bcl-2&\citet{tutusaus2018antiapoptotic}\\ \hline
Lapatinib&\makecell{HER2}&
\citet{esteva2010molecular} %johnston2006lapatinib}
&TAE684&PTEN&--\\ \hline
Nilotinib&\makecell{STAT5}&\citet{warsch2011high}&TKI258&CD49b&--\\ \hline
Nutlin.3&Bcl-2&\citet{drakos2011activation}&Topotecan&--&--\\ \hline
PD-0325901&MEK1&\citet{henderson2010mek}&ZD-6474&c-Kit&\citet{yang2006zd6474}\\ \bottomrule
\end{tabular}}
\end{table*}

%Driven by the need for a better understanding 

%\subsubsection{Feature selection with PCS inference}

%Personalized medicine requires identifying genomic disease signature from a patient, then matching it with the most effective therapeutic intervention \citep{costello2014}. Although a lot of recent progress has been made on developing novel drug sensitivity prediction algorithms, there has been very sparse literature on understanding how these algorithms relate to specific disease signatures. 

Beyond predictability, the PCS framework emphasizes stability throughout the data science life cycle so as to reduce reliance on particular human judgment calls. Accordingly, 
%we exploit the observation that in terms of the WPC-index, the Lasso, Lasso (ESCV), and RF are within 0.011 of the Gaussian kernel ridge regression, the most predictive model, when trained on each molecular profile separately, and 
we leverage and quantify the stability of important features under numerous data and model perturbations in staDRIP as follows: for each of the 24 drugs separately,

\begin{enumerate}
    \item \textbf{Use predictability as reality check:} select a set $\mathcal{M}$ of models with high predictive accuracy across a variety of metrics on the validation data. 
    %Is "signature" defined? I don't know what it means. how the validation data is set
    %
    \vspace{-3mm}
    \item \textbf{Compute stability of predictive features across data perturbations:} for each model $M\in \mathcal{M}$, refit the model $M$ to $B$ bootstrap replicates of the data, and compute the stability score of each feature as the proportion of $B$ bootstrap samples where the feature is selected (details in Appendix~\ref{apd:stability_scores}). Let $F_M$ denote the subset of features with high stability scores (e.g., top 10).%
    \vspace{-3mm}
    \item \textbf{Select predictive features that are stable across model perturbations:} take the intersection $\cap_{M\in \mathcal{M}}F_M$ as the stable predictive -omic features across data and model perturbations.
\end{enumerate}
\vspace{-3mm}

In our work, we are primarily interested in identifying proteins and genes that are predictive of drug responses as many drugs are directly related to known proteins and genes. Hence, considering the five models trained on the RNAseq and protein data separately, we take $\mathcal{M} = \{\text{RF}, \text{Lasso (ESCV)}, \text{Elastic Net}\}$. Note that while kernel ridge has the highest accuracy, it is omitted from $\mathcal{M}$ since there is no straightforward, computationally efficient method to select features from kernel ridge to the best of our knowledge. We also omit the Lasso from $\mathcal{M}$ as it generally has the worst predictive accuracy. For each remaining model in $\mathcal{M}$, we then take $F_M$ to be the 10 features with the highest stability scores and list those genes and proteins in the top 10 most stable features across all three models in Table~\ref{tab:signatures} in the Appendix.
In Table~\ref{tab:pubmedliterature}, we provide our main evidence for the utility of staDRIP, listing the single most stable protein for each drug along with independent publications that support these findings.
%There are several proteins and genes which we found to be among the top 10 features for all three models. These features that were deemed most stable in all three models are listed in Table \ref{tab:signatures} in the Appendix. 
%There are less RNAseq signatures identified as most stable in all models because (1) the number of RNAseq features (5000) is much larger than that of protein features, which makes feature selection considerably more difficult, and also makes our cutoff (top 10) fairly conservative, (2) the average correlation between RNAseq features is higher than that between protein features, which also makes extracting stable features more difficult; 
%Among the list of overlapping stable features in Table \ref{tab:signatures}, we also list in Table \ref{tab:pubmedliterature} the one with the highest stability score ranking along with recent biomedical publications, supporting the association between the protein and the drug\footnote{Detailed discussion on the features in in Table \ref{tab:signatures}, as well as how the literature search is done, is provided in Appendix \ref{apd:discussion}.}. 
Specifically, of the 24 proteins identified as most stable by staDRIP, 18 have been associated with the drug sensitivity or identified as a known or possible drug target in prior preclinical studies. See Appendix \ref{apd:discussion} for details of this literature evidence. %evidence from the literature.

Now in contrast to staDRIP, which finds stable predictive features across models with similar predictive accuracies, \xl{previous state-of-the-art methods \citep{barretina2012cancer, jang2014systematic} use only an elastic net to identify predictive -omics features of drug responses.}
%are there more recent papers with better results
To compare staDRIP to this elastic net approach, we extract the proteins with the highest stability score for each drug when taking $\mathcal{M} = \{\text{Elastic Net}\}$. Repeating the same literature search procedure as we did for the proteins identified by staDRIP, we found only 14 of the 24 proteins identified by the elastic net are known from previous clinical studies (see Table~\ref{tab:pubmedliterature_enet} in the Appendix). Detailed comparisons of the results of our method, staDRIP, and that of the elastic net can be found in Appendix \ref{apd:discussion}.

%Did we find the same protein features?? Detailed comparisons of our results and theirs can be found in Appendix
% #1 features in table 7, plus irinotecan&SLFN, PD-0325901, topotecan
%We note that though \cite{barretina2012cancer} used bootstrap sampling to incorporate stability with respect to data perturbations, they relied on a single model (Elastic net) for feature selection whereas our proposed pipeline further considers the stability across multiple models with similar prediction accuracies. 
%But the key difference between their method and our PCS inference and is that the original CCLE paper \citeyear{barretina2012cancer} uses a particular model, namely Elastic net, for feature selection, 
%However, they used a particular model, namely Elastic net, for feature selection, while we consider stability of predictive features across various similarly performing models. 
%As such, we can filter out signatures that are important for a particular model, but not for other models with similar or higher predictive accuracy. 

%By building the stability principle into our inference framework, we achieve higher precision in selecting biologically meaningful signatures. 

%\vspace{-2mm}

\section{Conclusion} \label{sec:discussion}
%\vspace{-2mm}
%Based on the PCS framework, we emphasize the importance of predictability, (computability), and stability as a minimum requirements for extracting scientific knowledge.
%This is especially critical in the era of personalized medicine, where heterogeneity among cell lines, patients, and diseases is undeniable, and poor generalization can be detrimental. Having seen that BMTMKL, the winner of the NCI-DREAM challenge, suffered from poor generalization to the CCLE data, this gives us further reason to investigate the stability of machine learning-based drug response prediction models and disease biomarkers selection to various data and model perturbations. 
Rooted by the PCS framework, we emphasize the importance of predictability, (computability), and stability as minimum requirements for extracting scientific knowledge throughout the staDRIP pipeline. We show that, guided by good prediction performance, incorporating a number of stability checks and extracting the stable parts of top-performing models can help to avoid the poor generalization exhibited by existing methods and can successfully identify candidate therapeutic targets for future preclinical research. We also acknowledge that while many stability considerations are built into staDRIP, there are inevitably human judgment calls that still impact our analysis. 
\xl{For example, we make a number of judgement calls in the data preprocessing stage, which we detail in Appendix \ref{sec:data}. Additionally, many other reasonable models such as ridge regression and gradient boosting could be considered in the staDRIP pipeline.}
We thus provide transparent and extensive documentation \href{https://github.com/Yu-Group/staDRIP}{here} to justify these decisions using domain knowledge when possible.

\section{Acknowledgments}

The authors would like to thank Karl Kumbier for his helpful comments. TT acknowledges support from the NSF Graduate Research Fellowship Program DGE-1752814. BY acknowledges the support from ARO Army Research Office grant W911NF-17-1-0005, Office of Naval Research Grant N00014-17-1-2176, the Center for Science of Information (CSoI), a US NSF Science and Technology Center, under grant agreement CCF-0939370, UCSF fund N7251, National Science Foundation grants NSF-DMS-1613002, 1953191, and IIS 1741340. BY is a Chan Zuckerberg Biohub investigator. This work was in part supported by the Mayo Clinic Center for Individualized Medicine.

\bibliography{bibliography}

\appendix

\section{Data} \label{sec:data}

To begin building the personalized drug response models, we leverage data from a panel of 397 human cancer cell lines that have both high-throughput molecular profiling and pharmacological data for 24 anticancer drugs from the Cancer Cell Line Encyclopedia (CCLE) project \citep{barretina2012cancer}. Specifically, -omics data from the CCLE was downloaded from DepMap Public 18Q3 (https://depmap.org/portal/download/). These cell lines encompass 23 different tumor sites and have been profiled for gene expression, microRNA expression, DNA methylation, and protein expression. Note that though the CCLE contains data from 947 cell lines, only 397 of these cell lines had data from all four molecular profiles of interest and pharmacological profiling. 

In addition to the molecular profiles, we obtained pharmacological profiling of 24 chemotherapy and target therapy drugs from the CCLE \citep{barretina2012cancer}. For each cell line-drug combination, the CCLE incorporated a systematic framework to measure molecular correlates of pharmacological sensitivity in vitro across eight dosages. We refer to \cite{barretina2012cancer} for details on this procedure, but given the fitted dose-response curves of growth inhibition from these experiments, we took the activity area, or AUC, to be the primary response of interest in this work. The AUC is defined as the area between the response curve and 0 (i.e., the no response reference level) and is a well-accepted measure of drug sensitivity \citep{jang2014systematic, barretina2012cancer}. In this case, the AUC is measured on an 8-point scale with 0 corresponding to an inactive compound and 8 corresponding to a compound with 100\% inhibition at all 8 dosages.

\begin{figure*}
    \centering
    \includegraphics[width = .9\linewidth]{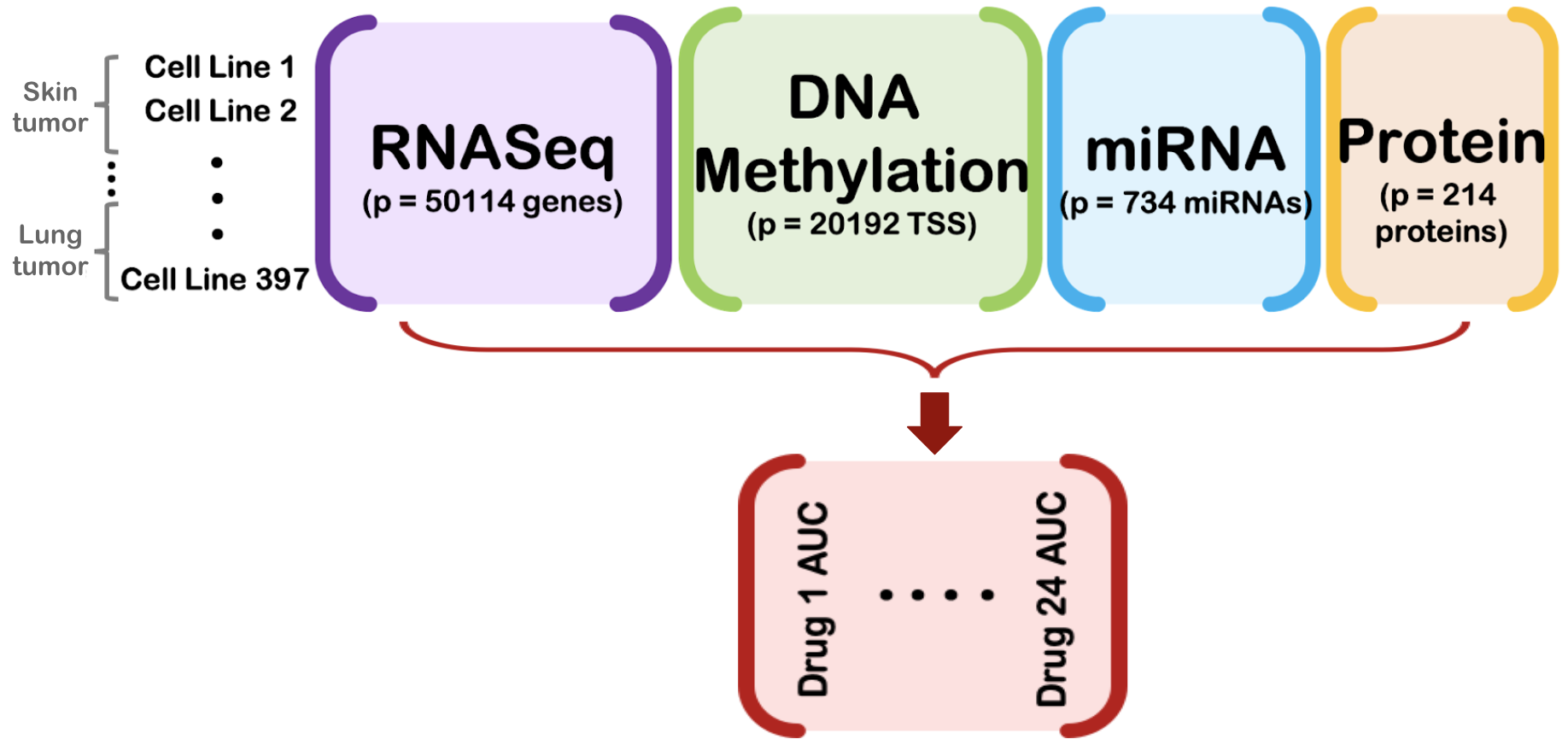}
    \caption{A graphical overview of the raw CCLE molecular profiling data sets, which are used to predict the drug responses of 24 therapeutic drugs, as measured via the drug response AUC.}
    \label{fig:data}
\end{figure*}

In Figure \ref{fig:data}, we provide a graphical summary of the raw molecular and pharmacological profiling data sets.

%For each of eight different dosages, two groups of cancer cells are cultured, and one group (the treatment group) is treated with drug at the given dose while the other group (the control group) is treated with blank culture medium. After 72 to 84 hours, assays were used to calculate the percent of growth inhibition by the drug-treated group relative to the control. A very potent drug can inhibit cancer cell growth with very low dose, and an ineffective drug may never reach certain percentage of grow inhibition (i.e. 50\%) even with a much higher dose.

%This dose-response data was then taken and fitted to one of three models - a constant model, a logistic (sigmoid) model, or a non-parametric spline interpolation (note that this last model represents less than 5\% of models). Using a Chi-squared test, the best model for the dose-response curve was selected, and the area between the response curve and 0 (i.e., the no response reference level) was defined to be the activity area, or AUC (see Figure 2.1). This AUC is measured on an 8-point scale with 0 corresponding to an inactive compound and 8 corresponding to a compound with 100\% inhibition at all 8 dosages. The AUC for these dose-response curves of growth inhibition will be the primary response of interest in this work.

\subsection{Data preprocessing} \label{sec:preprocess}

\begin{figure*}
    \centering
    \includegraphics[width = \linewidth]{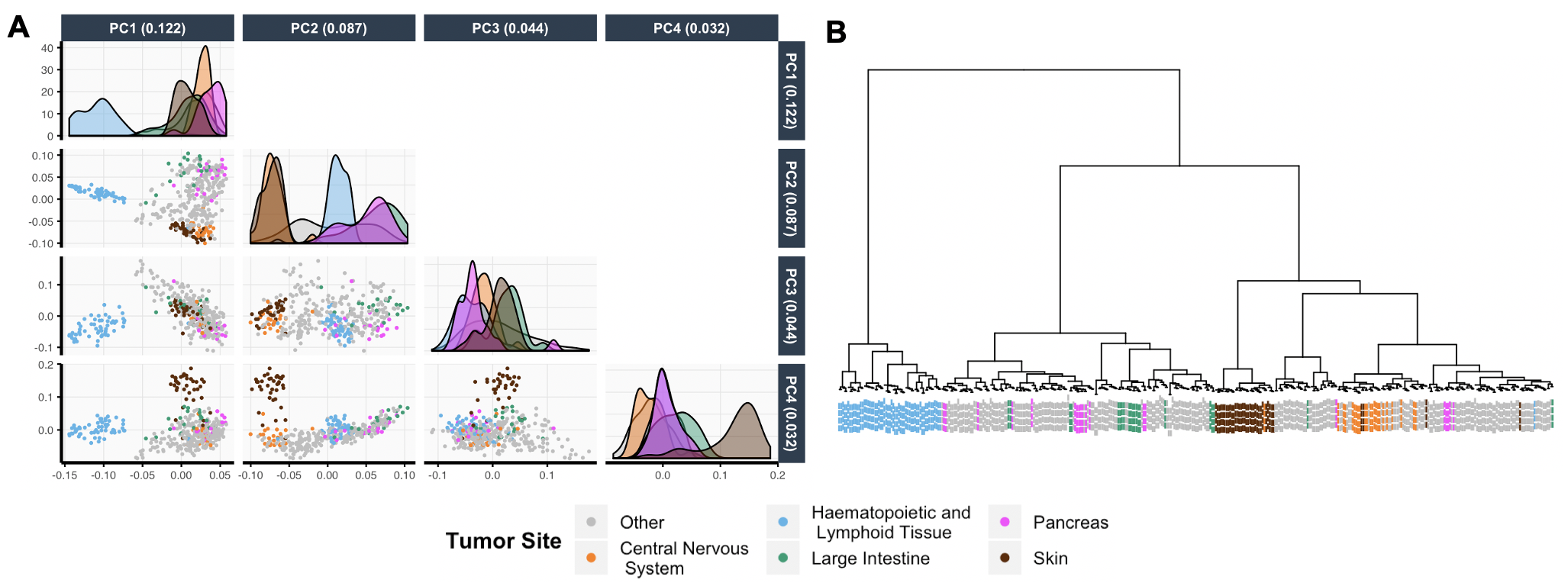}
    \caption{We apply (A) PCA and (B) hierarchical clustering (with Ward's linkage) to the log-transformed RNASeq data set and color the samples by their tumor site. For simplicity, we use color to distinguish between five prominent tumor sites and show the remaining tumor sites in grey. We also show the proportion of variance explained by each principal component in the subplot titles of (A). In both the PC plots and the hierarchical clustering dendrogram, we can see clusters of tumor sites, illustrating the inherent differences between tumor sites.}
    \label{fig:tissue}
\end{figure*}

Given the raw data described above, there are a couple areas of initial concern that warrant preprocessing. First, the cancer cell lines encompass 23 different tumor sites, and cell lines from the same tumor site tend to have more similar expression profiles than cell lines from different sites. To illustrate, we observe clusters of cell lines by tumor site when performing both hierarchical clustering and PCA on the RNASeq profile in Figure \ref{fig:tissue}. Due to these inherent differences between tumors, we chose to omit the cell lines from tumor sites with $<8$ cell lines. This reduces our sample size to 370 cell lines from 16 tumor sites. Here, we chose the threshold $8$ to ensure we have at least 2 cell lines from each tumor site in each of the training, validation, and test splits (using a 50-25-25\% partitioning scheme).

\iffalse
\begin{table}[h!]

\caption{Frequency of cell lines from each tumor site}
\label{tab:tissues}
\centering
\begin{tabularx}{8cm}{Xc}

%\begin{tabular}[t]{rc}
\toprule
\textbf{Tumor Site} & \textbf{\# of Cell Lines} \\
\midrule
Lung & 72\\
Haematopoietic and lymphoid tissue & 58\\
Skin & 36\\
Breast & 26\\
Central nervous system & 23\\
Ovary & 22\\
Large intestine & 21\\
Pancreas & 20\\
Endometrium & 15\\
Stomach & 14\\
Oesophagus & 13\\
Liver & 12\\
Urinary tract & 12\\
Autonomic ganglia & 9\\
Soft tissue & 9\\
Bone & 8\\
Kidney & 7\\
Upper aerodigestive tract & 6\\
Thyroid & 5\\
Pleura & 4\\
Prostate & 3\\
Biliary tract & 1\\
Salivary gland & 1\\
\bottomrule
%\end{tabular}
\end{tabularx}
\end{table}
\fi

In addition to reducing the number of samples in our analysis, we reduced the number of features to more manageable sizes before continuing with our analyses.  Originally, the molecular profiling data consisted of 734 miRNAs, 50114 genes, 20192 TSS, and 214 proteins. With only 370 cell lines, we aggressively preprocessed the number of genes and TSS by taking the top 10\% of genes (or 5000 genes) and top 20\% of TSS (or 4000 TSS) with the highest variance. We also transformed the miRNA and RNASeq expression values using the log-transformation log(x+1) in order to mitigate potential problems with highly skewed positive count values.

We recognize however that there were many other reasonable ways to preprocess this data. For instance, we could have taken the top 20\% of genes and top 40\% of TSS with the highest variance. Another common alternative would have been to filter features using marginal correlations with the response or using a multivariate prediction model (e.g., the Lasso). To assess robustness to these choices, we reran our prediction analysis using these alternative preprocessing procedures and saw that the prediction accuracies are higher using the variance-filtering preprocessing pipelines, as compared to the correlation-filtering and Lasso-filtering pipelines (see PCS \href{https://github.com/Yu-Group/staDRIP}{documentation}). Further, the smaller variance-filtered model gives similar prediction accuracies as the larger variance-filtered model. Thus, for simplicity moving forward, we use and focus primarily on the initially proposed variance-filtering procedure as it is less computationally expensive than the model with twice as many features and maintains similarly high accuracy.

\begin{figure*}[!ht]
    \centering
    \includegraphics[width = \linewidth]{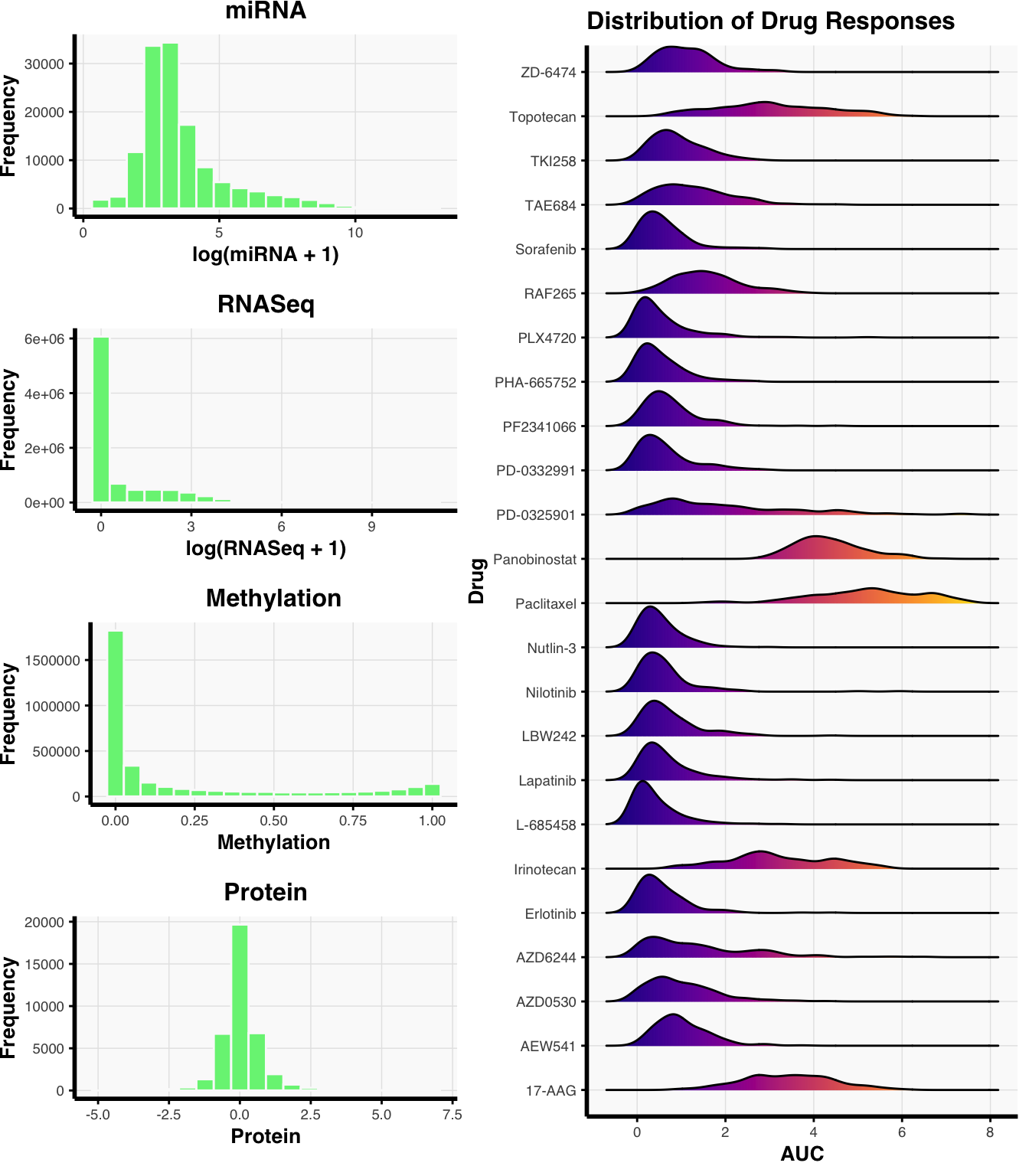}
    \caption{Left: Distribution of features in each of the four molecular profiles. Right: Distribution of the drug responses for each of the 24 drugs.}
    \label{fig:dist}
\end{figure*}

To summarize, after this preprocessing, we have 370 cell lines with data across the four molecular profiles of interest with 734 miRNAs (log-transformed), 5000 genes (log-transformed), 4000 TSS, and 214 proteins and pharmacological data, measured via the AUC drug response scores, for 24 anticancer compounds. We provide a visual summary of the preprocessed data and plot the overall distribution of features in the four molecular profiles as well as the distribution of the 24 drug responses in Figure \ref{fig:dist}.

\section{Prediction Models} \label{sec:prediction}

Like in data preprocessing, human judgment calls play a significant role in the modeling stage, including the decision of which methods to fit. Ideally, the chosen methods should have some justified connection to the biological problem at hand, but in our case, it is unclear which models or assumptions best fit the biological drug response mechanism a priori. Nevertheless, we have reasons to believe that the Lasso, elastic net, RF, and kernel ridge regression are particularly appealing fits for this problem. 

First, the Lasso assumes a sparse linear model, meaning that the effect of each feature is additive and only a sparse number of the features contribute to the drug sensitivity. The simplicity and interpretability of the Lasso makes it a popular tool for bioinformatics prediction tasks, so we choose to use the Lasso as a baseline model for our analysis. The elastic net is perhaps even more popular than the Lasso in drug response prediction studies \citep{jang2014systematic, barretina2012cancer}. Similar to the Lasso, the elastic net assumes linearity and some sparsity but is also able to better handle correlated features.  Beyond linearity, kernel ridge regression with a Gaussian kernel allows for more flexible, but less interpretable, functional relationships that are not necessarily linear. Kernel methods have been applied in previous case studies with great success \citep{costello2014} and are hence promising candidates for our study as well. Lastly, random forest can be viewed as a collection of localized, non-linear thresholded decisions rules (like on-off switches), which are well-suited for many biological processes that match the combinatorial thresholding (or switch-like) behavior of decision trees \citep{nelson2008lehninger}. Random forests are also invariant to the scale of the data. This is especially advantageous for integrating different data sets with varying scales and domain types (e.g., count-valued RNASeq expression, proportion-valued methylation data, continuous-valued protein expression).

In addition to fitting the aforementioned methods on each of the molecular profiles separately, we also tried fitting various data integration methods since incorporating multiple sources of -omics data can sometimes result in more accurate predictions than models built using only a single -omics sources \citep{costello2014, guvencc2019improving, simidjievski2019variational}. The most natural integration idea is to concatenate the -omics data sets together and to fit a single model (e.g., the elastic net) on the concatenated data. When fitting models like the Lasso, elastic net, and kernel ridge regression which are not scale-invariant, the molecular profiles are scaled to have columns with mean $0$ and variance $1$ to allow for fair comparisons between molecular types. We refer to this method as the concatenated data approach and use this as a baseline for evaluating data integration methods. More sophisticated methodology has also been proposed to integrate -omics data, including recent work using the X-VAE, a variational autoencoder for cancer data integration \citep{simidjievski2019variational}, and the BMTMKL, a Bayesian multitask multiple kernel learning method which won the NCI-DREAM 7 challenge \citep{costello2014}.

Note that though an alternative approach would have been to develop new methodology, we instead leverage these existing machine learning methods that have been rigorously vetted and have been shown to work well in many related problems. In fact, by examining the stable properties across these existing methods, we obtain high-quality scientific findings, as made evident by the abundance of supporting literature (see Table~\ref{tab:pubmedliterature}).

%We choose to focus our investigation on the above machine learning methods as they have been rigorously vetted in previous works and are known to work well in a plethora of related problems. Though an alternative approach would have been to develop new methodology, we instead leverage these existing well-vetted machine learning methods, show that many of them perform similarly well with regards to prediction, and extract the features that are deemed important and stable across several of these top-performing predictive models. In this work, we emphasize the importance of model interpretation and view prediction accuracy, not as an iron rod, but as a reality check to ensure that the model satisfactorily fits the data \citep{yu2020veridical}.

\subsection{Model hyperparameters}

To select hyperparameters in each of these methods, we use 5-fold cross validation, where the folds are stratified by tumor type. We also investigate using the estimation stability cross validation (ESCV) metric for selecting the Lasso's hyperparameter. This ESCV metric combines a stability measure within the cross-validation framework to yield more stable estimation properties with minimal loss of accuracy when using the Lasso \citep{lim2016estimation}.

For the X-VAE model, we adapt an X-shaped network architecture to train a variation autoencoder that learns joint representation of the RNAseq and protein data. In particular, we take the 2,000 RNAseq features with highest variance, since the number of cell lines is too small compared with the original number of RNAseq features. In our experiment, both the encoder and the encoder have one hidden layer. There are 128 neurons corresponding to the RNAseq protein in the hidden layer of the encoder and the decoder, and 32 neurons corresponding to the protein features. The latent representation has a dimension of 32. The dimension of the hidden layers and the latent representation are based on the recommendation of \citep{simidjievski2019variational}, and are not tuned. We used ELU activation and employed batch normalization and a dropout component with rate 0.2, as recommended by \citep{simidjievski2019variational}. The models were trained for 500 epochs using an Adam optimizer with a learning rate of 0.001. 

\subsection{Evaluation metrics} \label{sec:evaluation}
We primarily consider two evaluation metrics for prediction accuracy as each captures a different aspect of prediction - 1) $R^2$ value and 2) probabilistic concordance-index (PC-index). $R^2$ is defined as $1 - \frac{\text{MSE}(Y, \hat{Y})}{\text{Var}(Y)}$, where $\text{Var}(Y)$ denotes the variance of the observed responses, and $\text{MSE}(Y, \hat{Y})$ denotes the mean sum of squared errors between the predicted responses $\hat{Y}$ and observed responses $Y$. $R^2$ is a rescaling of the MSE that accounts for the amount of variation in the observed response and thus allows us to easily compare accuracies between drug response models with different amounts of variation in the observed response, but as with the MSE, $R^2$ can be heavily influenced by outliers.  PC-index is a measure of how well the predicted rankings agree with the true responses. This metric takes into account the variance of the drug responses but it also assumes that the drug responses follow a Gaussian distribution, which may not be true in some cases. We consider this metric because it is the primary method of evaluation in the NCI-DREAM 7 competition \citep{costello2014}. Given the large scale and breadth of this challenge, we compare our results to this work. For further details on the PC-index, we refer to \cite{costello2014}.

%$R^2$ is defined as $1 - \frac{\text{MSE}(Y, \hat{Y})}{\text{Var}(Y)}$, where $\text{Var}(Y)$ denotes the variance of the observed responses, and $\text{MSE}(Y, \hat{Y})$ denotes the mean sum of squared errors between the predicted responses $\hat{Y}$ and observed responses $Y$. $R^2$ is a rescaling of the MSE that accounts for the amount of variation in the observed response. $R^2$ thus allows us to easily compare accuracies between drug response models with different amounts of variation in the observed response, but as with the MSE, $R^2$ can be heavily influenced by outliers. 

In each of the evaluation metrics above, we receive a separate score for each of the 24 drug response models. It may also be beneficial to aggregate the 24 scores into a single number for concrete evaluation. In particular, \cite{costello2014} used a weighted average of the PC-indices to compare various models and referred to this evaluation metric as the weighted PC-index (WPC-index). To compare our results with the benchmark in \cite{costello2014}, we also consider the WPC-index in evaluating our models.

\subsection{Prediction results}\label{apd:pred_res}

\begin{table*}[h]

\caption{\label{tab:pred-res} Validation WPC-index and average $R^2$ across all 24 drug response models for various methods trained on each molecular profile separately and together. Higher values of $R^2$ and WPC-index indicate better fits.}
\centering
\scriptsize
\vspace{-3mm}
\setlength{\tabcolsep}{3pt}
\begin{tabular}[t]{@{}lccccc|ccccc@{}}
\toprule
\multicolumn{1}{c}{} & \multicolumn{5}{c|}{\textbf{Validation Set WPC-Index}} & \multicolumn{5}{c}{\textbf{Validation Set $\boldsymbol{R^2}$}} \\
\midrule
  & Methyl. & miRNA & Protein & RNASeq & Integrated & Methyl. & miRNA & Protein & RNASeq & Integrated \\
\midrule
Kernel Ridge &  0.600 & 0.603 & 0.617 & \textbf{0.631} & 0.624 & 0.111 & 0.104 & 0.168 & \textbf{0.231} & 0.200 \\
Elastic Net & 0.602 & 0.606 & 0.608 & 0.626 & 0.625 & 0.102 & 0.124 & 0.126 & 0.183 & 0.162 \\
Lasso & 0.597 & 0.605 & 0.609 & 0.620 & 0.620 & 0.117 & 0.105 & 0.121 & 0.172 & 0.176 \\
Lasso (ESCV) & 0.600 & 0.601 & 0.609 & 0.623 & 0.618 & 0.114 & 0.113 & 0.129 & 0.195 & 0.141 \\
RF & 0.599 & 0.594 & 0.606 & 0.626 & 0.622 & 0.124 & 0.088 & 0.123 & 0.214 & 0.196 \\
X-VAE & -- & -- & -- & -- & 0.617 & -- & -- & -- & -- & 0.188 \\
BMTMKL & -- & -- & -- & -- & 0.613 & -- & -- & -- & -- & 0.179 \\
\bottomrule
\end{tabular}

\end{table*}
\vspace{-.5mm}

In Table~\ref{tab:pred-res}, for various methods, we summarize the validation accuracy across all 24 drugs as measured by the average $R^2$ value and WPC-index. In Tables~\ref{tab:pred-res-best} and ~\ref{tab:test-err-drug}, we provide additional insights into the drug response prediction accuracies at the individual drug level. In Table~\ref{tab:pred-res-best}, we see that the best model depends on the particular drug, but the kernel ridge regression model works best on average. In Table~\ref{tab:test-err-drug}, we show the test errors from the kernel ridge regression fit for each drug separately.

\begin{table*}[h]
\centering
\caption{\label{tab:pred-res-best} For each molecular profile (or the integrated profile) used for training, we count the number of drugs (out of 24) for which each method performed the best and gave the highest validation $R^2$ compared to its six other competitors.}
\small
% \vspace{-1.5mm}
% \setlength{\tabcolsep}{3pt}
\begin{tabular}[t]{@{}lccccc@{}}
\toprule
% \multicolumn{6}{c}{\textbf{Validation Set WPC-Index}} \\
% \midrule
  & Methyl. & miRNA & Protein & RNASeq & Integrated \\
\midrule
Kernel Ridge & 7 & 5 & 16 & 12 & 6 \\
RF & 9 & 5 & 2 & 6 & 5 \\
Elastic Net & 1 & 8 & 1 & 1 & 2 \\
Lasso (ESCV) & 4 & 4 & 3 & 4 & 0 \\
Lasso & 3 & 2 & 2 & 1 & 5 \\
X-VAE & -- & -- & -- & -- & 2 \\
BMTMKL & -- & -- & -- & -- & 2 \\
\bottomrule
\end{tabular}

\end{table*}

\begin{table}[h]
\centering
\caption{\label{tab:test-err-drug} Test error for each drug using the RNASeq-based kernel ridge regression model}
\small
\begin{tabular}{ccc}
\toprule
\textbf{Drug}  & $\boldsymbol{R^2}$ & \textbf{PC-Index}\\
\midrule
17-AAG & 0.000 & 0.574\\
AEW541 & 0.034 & 0.558\\
AZD0530 & 0.037 & 0.560\\
AZD6244 & 0.425 & 0.675\\
Erlotinib & 0.254 & 0.615\\
Irinotecan & 0.307 & 0.644\\
L-685458 & 0.210 & 0.624\\
LBW242 & -0.001 & 0.511\\
Lapatinib & 0.208 & 0.607\\
Nilotinib & 0.258 & 0.590\\
Nutlin-3 & 0.022 & 0.549\\
PD-0325901 & 0.543 & 0.701\\
PD-0332991 & 0.218 & 0.596\\
PF2341066 & 0.091 & 0.564\\
PHA-665752 & 0.115 & 0.559\\
PLX4720 & 0.305 & 0.585\\
Paclitaxel & 0.369 & 0.670\\
Panobinostat & 0.446 & 0.679\\
RAF265 & 0.215 & 0.625\\
Sorafenib & 0.242 & 0.567\\
TAE684 & 0.024 & 0.576\\
TKI258 & 0.183 & 0.585\\
Topotecan & 0.240 & 0.630\\
ZD-6474 & 0.155 & 0.591\\
\bottomrule
\end{tabular}
\vspace{-2mm}
\end{table}

\section{PCS Inference}\label{apd:pcs_inference}

\subsection{Detailed description on the computation of stability scores}
\label{apd:stability_scores}

We next describe in detail how to compute the stability scores in the PCS-driven disease signature identification pipeline. Note that the following procedure is repeated for each of the 24 drugs.

%First, based on prediction accuracy, we select the following 3 models for interpretation: (1) Lasso with tuning parameter selected by ESCV (2) Lasso with tuning parameter selected by ESCV and (3) Random Forests. These three models were shown to have high predictive accuracies in Section~\ref{sec:pred_results} and hence are reasonable fits for the data. %??pls be consistent with forestS? how does RFs select features? do we mean iRF??) 
%Note that although Gaussian kernel ridge regression has the highest average prediction accuracy in terms of all three metrics (MSE, Pearson correlation, and WPC index), extracting relevant feature importance measures from the fitted kernel regression model remains an open question. Recent work by \citep{hainmueller2014kernel} proposed to estimate the marginal effect of each feature using partial derivatives of the response function. However, it is unclear whether these partial derivatives can be used for feature selection.

We randomly draw $B=100$ bootstrap samples $\mathcal{D}^{(b)}, b=1, \dots, 100$ from 
the training data. %For the CCLE data, bootstraps are appropriate data perturbations \citep{barretina2012cancer}. 
Then, for each bootstrap sample, we fit (1) an elastic net with tuning parameter selected by CV (2) a Lasso with tuning parameter selected by ESCV and (3) a random forest. %two-stage block-wise approach as described above, where we first fit a model using the protein data and then fit the residuals using RNAseq data.
Each model is fitted using the protein and RNAseq data separately since the integration approaches did not improve the prediction accuracy over simply using the RNASeq data only (see Table~\ref{tab:pred-res}). 
Next, for each feature $X_j$ from either the protein or RNAseq data set, let $\omega_j^{(b)}$ be defined in the following way: for the Lasso and elastic net, $\omega_j^{(b)}=1$ if the coefficient of $X_j$ is non-zero, and $\omega_j^{(b)}=0$ otherwise; for the random forest, $\omega_j^{(b)}$ is the MDI feature importance of $X_j$.
We then define the stability score $\text{sta}(X_j)$ of each feature $X_j$ as 
$\text{sta}(X_j) = \frac{1}{B}\sum_{b=1}^{B}\omega_j^{(b)}$ and rank the proteins and genes separately by the stability scores of the features. 
%The above procedure is repeated for each of the 24 drugs.

%Then, for each of the above three models 
%and for each of the 24 drugs, 
%we extract the ten proteins with the highest stability score among the protein features, and the ten genes with the highest stability score among the RNAseq features. Finally, we report the proteins and genes that are deemed the top 10 most stable in all three models.

%The average rank correlations between the stability scores of different models across all 24 drugs are given in Table \ref{tab:rankcorrelations}. Although these stability scores are positively correlated, the rank correlations are much smaller than 1, demonstrating that feature selection results are sensitive to modeling decisions. One potential reason for these low correlations is that the top features selected by different models are different but belong to the same biological pathway. However, this hypothesis is hard to verify using the available data and known predictors of drug response.

\iffalse

\begin{table*}
\caption{Rank correlations between stability scores from different models.\label{tab:rankcorrelations}}
\centering
\begin{tabular}[t]{ccc}
\toprule
\textbf{} & \textbf{Protein} &  \textbf{RNAseq}\\ \midrule
Lasso-ESCV and Lasso-CV & 0.44 & 0.41 \\ 
Lasso-ESCV and RF & 0.50 & 0.29\\
Lasso-CV and RF & 0.28 & 0.22\\
\bottomrule
\end{tabular}
\end{table*}

\fi

\begin{table*}[ht]
\caption{Stable protein and RNAseq signatures. A feature is included if it is among the top 10 most stable features under 3 different machine learning models (i.e., elastic net, Lasso (ESCV), and random forests). The stability of the features are computed from the PCS inference framework in staDRIP. Blank cells indicate that no features appeared among the top 10 most stable features for all three models.\label{tab:signatures}}
\centering
\scriptsize
\begin{tabular}{ccc}\toprule
\thead{Drug name}&\thead{Protein signature}&\thead{RNAseq Signature}\\\midrule
17-AAG &Bax, p53, Caspase-7, eIF4E &CTD, AP2S1, BZW2\\ \hline
\makecell{AEW541}&\makecell{Akt, Smad1, p27, PTEN, RAD51}&B4GALT3, SEMA3B\\ \hline
AZD0530&p38, c-Kit
&HPGD\\ \hline
AZD6244&\makecell{PI3K-p85, TFRC, Bax}&SPRY2, RP11, LYZ, DUSP6, PRSS57\\ \hline
Erlotinib&EGFR, Beclin, P-Cadherin&PIP4K2C, SEC61G\\ \hline
Irinotecan&\makecell{MDMX\_MDM4, Src}&\\ \hline
L-685458&YAP, VEGFR2, Src&\\ \hline
LBW242&ASNS&MRPL24\\ \hline
Lapatinib&\makecell{HER2, HER3, EGFR, Rab25, Heregulin}&STARD3\\ \hline
Nilotinib&\makecell{STAT5, c-Kit, SHP-2, Src, p27}&\\ \hline
Nutlin.3&Bcl-2, Bax&\\ \hline
PD-0325901&MEK1, Bax, TFRC, PI3k
&SPRY2, DUSP6, ETV4\\ \hline
PD-0332991&\makecell{Bcl-2, MDMX\_MDM4, Src}&\\ \hline
PF2341066&c-Met
&CAPZA2\\ \hline
PHA-665752&MEK1, c-Met&FMNL1\\ \hline
PLX4720&\makecell{MEK1, Bax, PREX1, Beclin}&FABP7\\ \hline
Paclitaxel&Src, beta-Catenin&ORMDL2\\ \hline
Panobinostat&VEGFR2, Src&\\ \hline
RAF265&PI3K-p85, FOXO3a, eEF2K  
&RETN\\ \hline
Sorafenib&Bcl-2, Src&\\ \hline
TAE684&PTEN, Akt, p70S6K, Bcl-2
&H1FX\\ \hline
TKI258&CD49b, C-Raf
&\\ \hline
Topotecan&c-Met&OSGIN1\\ \hline
ZD.6474&c-Kit, STAT5-alpha&\\ \bottomrule
\end{tabular}

\end{table*}

\subsection{Discussion on the disease signatures identified by the PCS pipeline}
\label{apd:discussion}

In Table \ref{tab:signatures}, we list the proteins and genes which we found to be stable and among the top 10 features for all three methods. Among these stable features, we list them in decreasing order by the sum of stability score rankings. 
Though we identify fewer stable genes, this is most likely due to two reasons. First, there are 5000 genes in the model, compared to only 214 proteins, so thresholding at the top 10 genes is extremely conservative. Secondly, the average correlation between genes is higher than that between proteins, adding to the instability.

With regards to the identified protein signatures, we can roughly classify them into three categories. The first category contains those that are known targets of the corresponding target therapy drugs. For example, Erlotinib is a medication used to treat non-small cell lung cancer (NSCLC) and pancreatic cancer. It is an EGFR inhibitor and is specifically used for NSCLC patients with tumors positive for EGFR exon 19 deletions (del19) or exon 21 (L858R) substitution mutations. Correspondingly, EGFR is ranked in the top ten stable proteins in all three models. Other such examples include the drug Lapatinib and its target HER2, PD-0325901 and its target MEK, PHA-665752 and its target c-Met, and ZD-6474 and its target c-Kit. 

The second category contains those that are not known to be direct targets of the drug but have been shown in preclinical studies to be potential therapeutic targets or are associated with drug resistance. For example, \cite{ling2014fl118} identified a potential application of the drug Irinotecan as an MdmX inhibitor for targeted therapies, and in our pipeline, MdmX had the highest stability score for all three models. As another instance, we identified MEK1 as a top protein signature, ranked by stability score, for the drug PLX4720 while \cite{emery2009mek1} showed that MEK1 mutations confer resistance to PLX4720. 

%Compare this with the original CCLE paper \citep{barretina2012cancer} , which uses Elastic net to identify predictors of drug sensitivity. Of the 24 top-1 signatures (one for each drug) identified in \citep{barretina2012cancer}, only 9 of them are known predictors of drug sensitivity (see Figure 8 and Table 7 in the supplementary material of \citep{barretina2012cancer} for these signatures).
% #1 features in table 7, plus irinotecan&SLFN, PD-0325901, topotecan We note that \citep{barretina2012cancer} also uses bootstrap sampling to measure stability with respect to data perturbation. But the key difference between their method and our PCS inference and is that the original CCLE paper \citep{barretina2012cancer} uses a particular model, namely Elastic net, for feature selection, while we consider stability of predictive features across various similarly performing models. As such, we can filter out signatures that are important for a particular model, but not for other models with similar or higher predictive accuracy. By building the stability principle into our inference framework, we achieve higher precision in selecting biologically meaningful signatures. 

\begin{table*}[ht]
\scriptsize
\centering
\caption{Most stable protein associated with each drug, as identified by the elastic net, along with preclinical evidence that supports the association between the listed protein and drug sensitivity. \label{tab:pubmedliterature_enet}}
\vspace{-2mm}
{\begin{tabular}{@{}ccc|ccc@{}}\toprule
\textbf{Drug} & \textbf{Protein} & \textbf{Supporting Literature} & \textbf{Drug} & \textbf{Protein} & \textbf{Supporting Literature} \\ \midrule
17-AAG &p53&\citet{naito2010promotion}&PD-0332991&\makecell{Bcl-xL}&\citet{chen2017dual}\\ \hline
%AEW541&\makecell{PTEN}&\makecell{\cite{patel2014pten}\\\cite{isebaert2011insulin}}\\ \hline
AEW541&Akt&\citet{attias2011insulin}&PF2341066&PEA15&--\\ \hline
AZD0530&p38
&\citet{yang2010effect}&PHA-665752&MEK1&--\\ \hline
AZD6244&\makecell{CD20}&--&PLX4720&\makecell{MEK1}&\citet{emery2009mek1}\\ \hline
Erlotinib&P-Cadherin&--&Paclitaxel&Src&\citet{le2011src}\\ \hline
Irinotecan&\makecell{RAD51}&\citet{shao2016gefitinib}&Panobinostat&Src&--\\ \hline
L-685458&VEGFR2&--&RAF265&PI3K-p85&\citet{mordant2010dependence}\\ \hline
LBW242&Caspase-7&--&Sorafenib&14-3-3 epsilon&\citet{wu2015involvement}\\ \hline
Lapatinib&\makecell{HER2}&
\citet{esteva2010molecular} %johnston2006lapatinib}
&TAE684&Akt&--\\ \hline
Nilotinib&\makecell{p27}&\citet{liu2011inhibition}&TKI258&14-3-3 epsilon&--\\ \hline
Nutlin.3&Bcl-2&\citet{drakos2011activation}&Topotecan&14-3-3 epsilon&--\\ \hline
PD-0325901&MEK1&\citet{henderson2010mek}&ZD-6474&c-Kit&\citet{nishioka2007zd6474}\\ \bottomrule
\end{tabular}}
\end{table*}

The third category are proteins that do not belong to the two categories above. Still, these biomarkers are predictive of the drug response under various model and data perturbations. Given the evidence in the scientific literature that supports many of our identified features, the proteins in this category may be potential candidates for future preclinical investigation.

Among the list of overlapping stable features in Table \ref{tab:signatures}, we list in Table \ref{tab:pubmedliterature} the one with the highest stability score ranking along with recent biomedical publications, supporting the association between the protein and the drug.
The procedure of this literature search is as follows: we first searched for papers where the protein and the drug co-occurs. Then for each paper, we read the introduction section to understand their main conclusions. 
Each of the 18 papers listed in Table \ref{tab:pubmedliterature} includes sentences such as ``our findings suggest that the over-expression of this protein will increase drug sensitivity/resistance" or ``this protein is a potential (or known) therapeutic target for the drug".
Out of the 24 predictive protein signatures that we identify as most stable, 18 of them have existing preclinical studies that confirm the effectiveness of our stability analysis. 

In Table \ref{tab:pubmedliterature_enet}, we list the protein with the highest stability score when fitting an elastic net to 100 bootstrap samples of the training data for each of the 24 drugs. This approach of finding predictive -omics features was previously used in \citep{barretina2012cancer, jang2014systematic}. Compared with staDRIP, which searches for stable features across different models, this approach only uses a single model (i.e, an elastic net) for feature selection. We repeat the same literature search procedure as we did for our findings and found that among the 24 most stable protein features identified by elastic net, only 14 are known from previous clinical studies. 
For 10 drugs, the most stable protein from elastic net and that from staDRIP is the same, and among these 10 proteins, 9 are implicated in the existing literature. For the other 14 drugs, 5 proteins identified by elastic net are implicated in the existing literature, while 9 protein features identified by staDRIP are implicated in the existing literature. 

%\section{Second Appendix}\label{apd:second}

%This is the second appendix.

\end{document}